\documentclass{physeauth}
\usepackage{graphicx}
\usepackage{amsmath}
\usepackage{amssymb}

\begin{document}

\begin{frontmatter}

\title{Probing e-e interactions in a periodic array of GaAs quantum wires}

\author{Y.~Jompol\thanksref{thank1}},
\author{C.~J.~B.~Ford},
\author{I.~Farrer},
\author{G.~A.~C.~Jones},
\author{D.~Anderson},
and
\author{D.~A.~Ritchie}

\address{Cavendish Laboratory, University of Cambridge, J J Thomson Avenue, Cambridge CB3 0HE, UK}
\author{T.~W.~Silk}
and
\author{A.~J.~Schofield}

\address{School of Physics and Astronomy, University of Birmingham, Edgbaston, Birmingham B15 2TT, UK}

%\address[address2]{Institute for Smart, Nowhere}

%\address[address3]{Famous Laboratory, Somewhere}

\thanks[thank1]{
Corresponding author.
E-mail: yj207@cam.ac.uk}

\begin{abstract}

We present the results of non-linear tunnelling spectroscopy between
an array of independent quantum wires and an adjacent
two-dimensional electron gas (2DEG) in a double-quantum-well
structure. The two layers are separately contacted using a
surface-gate scheme, and the wires are all very regular, with
dimensions chosen carefully so that there is minimal modulation of
the 2DEG by the gates defining the wires. We have mapped the
dispersion spectrum of the 1D wires down to the depletion of the
last 1D subband by measuring the conductance \emph{G} as a function
of the in-plane magnetic field \emph{B}, the interlayer bias $V_{\rm
dc}$ and the wire gate voltage. There is a strong suppression of
tunnelling at zero bias, with temperature and dc-bias dependences
consistent with power laws, as expected for a Tomonaga-Luttinger
Liquid caused by electron-electron interactions in the wires. In
addition, the current peaks fit the free-electron model quite well,
but with just one 1D subband there is extra structure that
may indicate interactions.
\end{abstract}
\end{frontmatter}

%[main text]
%\section{Introduction}
Electron-electron interactions in a one-dimensional (1D) metal are
predicted to cause the formation of a Tomonaga-Luttinger Liquid
(TLL) in which there are both elementary spin and charge
excitations. Other properties of the TLL have been studied in a
variety of systems (see e.g.\/ \cite{Bockrath}), but only a few have
attempted to detect the spin-charge separation directly, e.g. by
photoemission \cite{Kim} and tunnelling
\cite{Tserkovnyak,Auslaender} spectroscopy.

Following measurements of the spectral function of 1D wires by
Kardynal \textit{et al.},\cite{Kardynal} Altland \textit{et
al.}\cite{Altland} proposed their use in detecting the spin-charge
separation. The tunnelling current $I$ between the 1D wires and an
adjacent low-disorder 2D layer depends on the overlap between the
spectral functions of the two systems, which is varied by using an
in-plane magnetic field $B$ perpendicular to the wires to offset the
two in $k$-space; a bias $V_{\rm dc}$ between the layers is used to
investigate the energy dependence. In a non-interacting system,
peaks in the conductance $G$ follow the 1D and 2D subbands. In
contrast, for a TLL, there should be two features, for spin and
charge, instead of one for a non-interacting 1D
subband.\cite{Altland,Grigera}

We have previously used ion-beam lithography to make separate
contact to two such layers of electrons, for investigating arrays of
1D wires\cite{Macks} and antidots\cite{Zolleis}, but here we adopt a
simpler technique that uses just surface gates and allows
measurements even when the last 1D subband is nearly depleted. With
this technique, we probe the single-particle spectrum of the 1D
system down to single subband. We also find a zero-bias anomaly
(ZBA) in which the tunnelling current between a single 1D subband
and the 2D electron gas (2DEG) is strongly suppressed at low
temperatures.

\begin{figure}[t] %figure1
%h=here, t=top, b=bottom, p=separate figure page
\begin{center}\leavevmode
\includegraphics[width=1\linewidth]{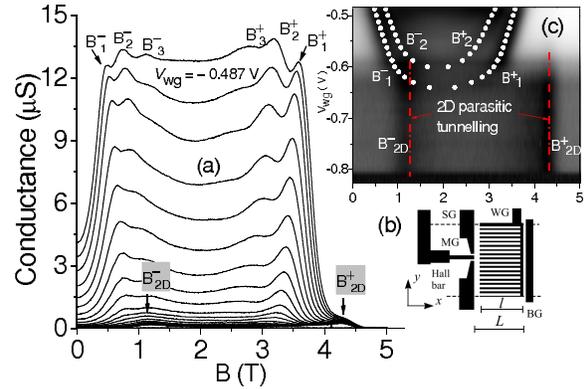}
\caption{(a) Tunnelling conductance $G$ showing resonant peaks in
the transverse magnetic field measured at 1.8 K. $V_{\rm wg}$ was
incremented from $-0.487$~V (top curve) to $-0.90$~V in steps of
12~mV. (b) Device layout; gates are shown in black. (c) Greyscale of
the data in (a), normalising each trace to the same maximum height.
Dark regions are peaks in $G$, with the white dots on top
%circles and crosses respectively
indicating the first and second 1D subbands, respectively. The 2D
parasitic tunnelling from region $h$ does not change with $V_{\rm
wg}$ -- its peaks stay at 1.27~T and 4.31~T, represented as dash
lines.} \label{figsubbands}\end{center}\end{figure}

%\section{Device fabrications and measurements}
The 1D system was formed as a array of long wires on a
double-quantum-well (DQW) structure. The wafer
% (C2617) was grown by molecular-beam epitaxy,
comprised two identical 18 nm GaAs quantum wells separated by a 14
nm Al$_{0.33}$Ga$_{0.67}$As tunnel barrier. 40 nm Si-dopant layers
($\sim$ $10^{18}$ cm$^{-3}$) above and below the two quantum wells
provide electron concentrations of $2.8~(1.64) \times 10^{11}$
cm$^{-2}$ with mobilities of about $8~(5) \times 10^{5}$ cm$^{2}$
V$^{-1}$ s$^{-1}$ in the top (bottom) well, as measured at 1.8 K.
The device was processed into a 100 $\mu$m-wide Hall bar.
Electron-beam lithography was used to define
% a split gate (SG) with a midline gate (MG), a depletion gate (BG), and
an array of $17.5 \times 0.1 \mu$m wire gates (WG) with a 270 nm
period, all the way across the mesa (see Fig.~\ref{figsubbands}(c)).

%\section{Results}
The two-terminal conductance $G = {\rm d}I/{\rm d}V_{\rm dc}$
\sloppy was measured at temperatures $T$ down to $<$100~mK
% in a dilution refrigerator
using an ac excitation voltage of 5~$\mu$V at 77~Hz added to a dc
voltage. Independent Ohmic contacts to the individual layers were
achieved by a surface-gate depletion scheme\cite{Nield}. Here, a
split gate SG, positioned at one end of the wires (see inset,
Fig.~1), was biased to $V_{\rm sg} = -2.25$~V, and a `mid-line' gate
(MG) was set to $V_{\rm mg} = +0.60$~V. This defined a conducting 1D
channel only in the top 2DEG. An additional depletion gate labelled
BG was biased as $V_{\rm bg} = -0.56$ V in order to isolate the top
2DEG from the drain Ohmic contact. Current was carried by electrons
tunnelling between the layers in the region of length $L$ between SG
and BG. A negative bias $V_{\rm wg}$ on the gate wires (of length
$l$) nearly covering the whole tunnelling area depleted the upper
2DEG into an array of 1D wires. However, a small ungated region $h$
of width $L-l = 0.9 \mu$m provides a current path to the entrances
to the 1D wires. The widths of $h$ and of the wire gates were chosen
carefully such that even with just a single 1D subband in the top
2DEG the wires remained conducting with minimal modulation of the
lower 2DEG. The long narrow 2D $h$ region inevitably contributes a
2D parallel tunnelling path that appears in the measurements.
However, this is small and independent of the tunnel current from
the wires.

The magnetic field $B$ shifts the origin of the 2D Fermi circle in
$k$-space relative to that of the 1D system by $k_{\rm
x}=eBd/\hbar$, where $d$ is the centre-to-centre tunnelling distance
between the two quantum wells and the $x$-axis is along the wires.
Tunnelling occurs at overlapping parts of the two spectral functions
and conserves the electron energy and momentum. By applying a
positive bias $V_{\rm dc}$ to the 2DEG, electrons tunnel into
excited states of the 2DEG, from matching states below the Fermi
energy in the 1D wires. At $V_{\rm dc} = 0$, there is a series of
peaks as a function of $B$ voltage. As shown in
Fig.~\ref{figsubbands}(a), there is a pair of peaks $B_i^\pm$ for
the  $i$-th 1D subband. Six peaks are observed at $V_{\rm wg} =
-0.487$~V (upper trace, Fig.~\ref{figsubbands}(a)), corresponding to
three occupied 1D subbands. With decreasing $V_{\rm wg}$, the 1D
subband spacing increases and the 1D density decreases. Hence the
number of subbands can be reduced to just one. At $V_{\rm wg} {\sim}
-0.65$~V the wires become insulating but $V_{\rm wg}$ is too small
to pinch off the bottom 2DEG. Some parasitic current can flow via
tunnelling in the parallel region $h$.
% A clearer picture of pinch off of the wires is obtained by
Normalising the conductance (Fig.~\ref{figsubbands}(b)) reveals the
depletion of each subband and the 2D-2D parasitic tunnelling peaks.
The bottom 2DEG becomes depleted at $V_{\rm wg} = -0.82$~V.

\begin{figure}[h] %figure2
%h=here, t=top, b=bottom, p=separate figure page
\begin{center}\leavevmode
\includegraphics[width=0.7\linewidth]{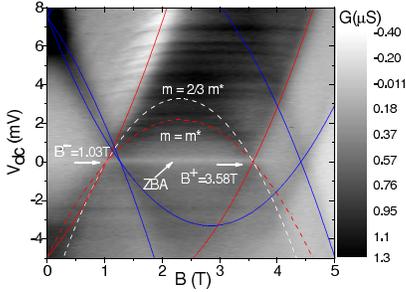}
\caption{Tunnelling conductance $G$ vs $V_{\rm dc}$ and $B$.}
\label{greyGvsBV}\end{center}\end{figure}

Fig.\/ \ref{greyGvsBV} shows $G$ as a function of the DC interlayer
bias ($V_{\rm dc}$) and $B$ as a greyscale plot at $V_{\rm wg} =
-0.62$~V, well into the region where there is just one 1D subband.
In the absence of interactions, there should be peaks in $G$ that
follow the 1D and 2D parabolic dispersion relations, shown as lines.
The parasitic 2D-2D tunnelling in the ungated $h$ region is also
shown, as thick blue lines that intersect B=0 at $+7.5$~meV, where
for 1D-2D tunnelling is occurred at $-5$~meV, as red. The crossing
points along $V_{\rm dc}$ = 0 are obtained at B$^{-}$ = 1.03 T and
B$^{+}$ = 3.58 T, giving the electron density in the wires $n_{\rm
1D} \simeq 40 \mu$m$^{-1}$. These two sets of dispersions can be
distinguished clearly and their singularities follow the lines well.
This is in good agreement with our calculated single-particle
results, such as the negative conductance at the top left where we
use the electron effective mass in GaAs (m$^{\ast}=0.067$m$_{\rm
e}$) and $d=32$~nm in our system. However, a lower effective mass
such as $2/3~m^{\ast}$ may fit the broad peak better; this could be
caused by interactions in the wires.

Moreover, a feature that cannot be explained in the noninteracting
model is the `zero-bias anomaly' (ZBA), the strong suppression of
conductance along $V_{\rm dc}=0$, visible as a bright line in Fig.\/
\ref{greyGvsBV}. This is likely to be related to the energy cost for
an electron to tunnel into or out of a 1D wire, as it disturbs the
line of electrons on either side of it. It has previously been
observed in 1D-1D tunnelling\cite{Tserkovnyak} and $G$ is expected
to have power-law dependences on $V_{\rm dc}$ and temperature $T$.

\begin{figure}[h] %figure3
%h=here, t=top, b=bottom, p=separate figure page
\begin{center}\leavevmode
\includegraphics[width=0.8\linewidth]{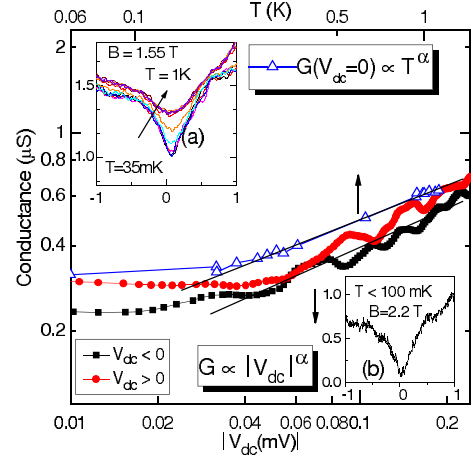}
\caption{Minima of the ZBA $vs$ $T$
%for $V_{\rm wg} = -0.61$ V
for $B =1.55$~T ($0.3 \mu$S was subtracted to allow for the
parasitic current) and $G(V_{\rm dc})$ for $B=2.2$~T. Inset (a): ZBA
at various $T$ up to 1.2~K at $B=1.55$~T. Inset (b): $G(V_{\rm dc})$
on a linear scale.
% The slope of $G(T)$ and $G(V_{\rm dc})$ is found to
% be 0.41 and 0.45, respectively.
} \label{ZBA}\end{center}\end{figure}

Fig. \ref{ZBA} shows the ZBA as a function of $V_{\rm dc}$ at
various temperatures (inset (a)) close to $B=B_1^-$, and midway
between $B_1^-$ and $B_1^+$ (where the parasitic conductance is
almost negligible) (inset (b)). The main graph shows the value of
the ZBA minimum as a function of $T$ as in inset (a) and of $V_{\rm
dc}$ (as in inset (b)), both on log-log axes. The power-law exponent
$\alpha$ found from the slopes of $G(T)$ and $G(V_{\rm dc})$ are
similar, $\alpha = 0.4 \pm 0.1$, and comparable to that found for
cleaved-edge overgrown quantum wires \cite{Tserkovnyak}.

In summary, we have measured momentum-resolved tunnelling from a 1D
electron system fabricated as an array of GaAs quantum wires, into a
high-mobility 2D electron gas. The number of occupied subbands can
be reduced from three to one using a gate voltage. For the last 1D
subband we have mapped the single-particle excitation spectrum of
the tunnelling conductance. A strong suppression of the tunnelling
current at zero bias (the zero-bias anomaly, ZBA) shows that
interactions are very important in these 1D wires. The extracted TLL
power-laws are consistent with wires made by
cleaved-edge overgrowth.\\

Y. Jompol acknowledges the Thai Ministry of Science Scholarships and
EPSRC for funding. 

\vspace{-0.5cm}

\begin{thebibliography}{10}
%\bibitem{Tarucha} S.~Tarucha, T. Honda, and T.~Saku, Solid State Commun.~{\bf 94} 413 (1995).
%\bibitem{Tarucha} S.~Tarucha, \emph{et al.\/}, Solid State Commun.~{\bf 94} 413 (1995).

%\bibitem{Bockrath} M.~Bockrath, D.H.~Cobden, Jia Lu, A.G.~Rinzler, R.E.~Smalley, L.~Balents, and P.L.~McEuen, Nature {\bf 397} 598 (1999).
\bibitem{Bockrath} M.~Bockrath, \emph{et al.\/}, Nature {\bf 397} 598 (1999).

%\bibitem{Kim} C.~Kim, A.Y.~Matsuura, Z.-X.~Shen, N.~Motoyama, H.~Eisaki, S.~Uchida, T.~Tohyama, and S.~Maekawa, Phys.~Rev.~Lett.~{\bf 77} 4054 (1996).
\bibitem{Kim} C.~Kim, \emph{et al.\/}, Phys.~Rev.~Lett.~{\bf 77} 4054 (1996).

%\bibitem{Tserkovnyak} Y.~Tserkovnyak, B.I.~Halperin, O.M.~Auslaender, and A.~Yacoby, Phys.~Rev.~B.~{\bf 68} 125312 (2003).
\bibitem{Tserkovnyak} Y.~Tserkovnyak, \emph{et al.\/}, Phys.~Rev.~B.~{\bf 68} 125312 (2003).

%\bibitem{Auslaender} O.M.~Auslaender, H.~Steinberg, A.~Yacoby, Y.~Tserkovnyak, B.I.~Halperin, K.W.~Baldwin, L.N.~Pfeiffer, and K.W.~West, Science {\bf 308} 88 (2005).
\bibitem{Auslaender} O.M.~Auslaender, \emph{et al.\/}, Science {\bf 308} 88 (2005).

%\bibitem{Kardynal} B.~Kardynal, {\em et al.\/}, Phys. Rev. Lett. {\bf 76}, 3802 (1996).
\bibitem{Kardynal} B.~Kardynal, {\em et al.\/}, Phys. Rev. Lett. {\bf 76}, 3802 (1996).

%\bibitem{Altland} A.~Altland, C.H.W.~Barnes, F.W.J.~Hekking, and A.J.~Schofield, Phys.~Rev.~Lett.~{\bf 83} 1203 (1999).
\bibitem{Altland} A.~Altland, \emph{et al.\/}, Phys.~Rev.~Lett.~{\bf 83} 1203 (1999).

%\bibitem{Grigera} S.A.~Grigera, A. J.~Schofield, S.~Rabello, and Q.~Si, Phys. Rev. B~{\bf 69} 245109 (2004).
\bibitem{Grigera} S.A.~Grigera, \emph{et al.\/}, Phys. Rev. B~{\bf 69} 245109 (2004).

%\bibitem{Macks} L.D.~Macks, C.H.W.~Barnes, J.T.~Nicholls, W.R.~Tribe, D.A.~Ritchie, P.D.~Rose, E.H.~Linfield and M.~Pepper, Physica E, {\bf 6}, 518 (2000).
\bibitem{Macks} L.D.~Macks, \emph{et al.\/}, Physica E, {\bf 6}, 518 (2000).

%\bibitem{Zolleis} K.R.~Zolleis, C.J.B.~Ford, B.~Kardynal, D.A.~Ritchie, E.H.~Linfield, P.D.~Rose and G.A.C.~Jones, Phys. Rev. Lett., {\bf 89}, 146803 (2002).
\bibitem{Zolleis} K.R.~Zolleis, \emph{et al.\/}, Phys. Rev. Lett., {\bf 89}, 146803 (2002).

%\bibitem{Nield} S.A.~Nield, J.T.~Nilcholls, W.R.~Tribe, M.Y.~Simmons, and D.A.~Ritche, J.~Appl.~Phys.~{\bf 87} 4036 (2000).
\bibitem{Nield} S.A.~Nield, \emph{et al.\/}, J.~Appl.~Phys.~{\bf 87} 4036 (2000).


%\bibitem{Levy} E.~Levy, A.~Tsukernik, M.~Karpovski, A.~Palevski, B.~Dwir, E.~Pelucchi, A.~Audra, E.~Kapon, and Y.~Oreg, Phys.~Rev.~Lett.~{\bf 97} 196802 (2006).
%\bibitem{Levy} E.~Levy, \emph{et al.\/}, Phys.~Rev.~Lett.~{\bf 97} 196802 (2006).

%\bibitem{Yacoby} A.~Yacoby, H.L.~St\"{o}rmer, Ned S.~Wingreen, L.N.~Pfeiffer, K.W.~Baldwin, and K.W.~West, Phys.~Rev.~Lett.~{\bf 77} 4612 (1996).
%\bibitem{Yacoby} A.~Yacoby, \emph{et al.\/}, Phys.~Rev.~Lett.~{\bf 77} 4612 (1996).

%\bibitem{Voit} ??J.~Voit, Phys.~Rev.~B.~{\bf 47} 6740 (1993).
%\bibitem{Voit} ??J.~Voit, Phys.~Rev.~B.~{\bf 47} 6740 (1993).
\end{thebibliography}
\end{document}